\documentclass[aps, prb, reprint, superscriptaddress, floatfix]{revtex4-2}
\usepackage{graphicx}% Include figure files
\usepackage{dcolumn}% Align table columns on decimal point
\usepackage{bm}% bold math
\usepackage{physics}
\usepackage{amsfonts}
\usepackage{mathrsfs}
\usepackage{subfigure}
\usepackage[svgnames]{xcolor}
\usepackage[colorlinks=true,linkcolor=NavyBlue,anchorcolor=red,citecolor=NavyBlue, urlcolor=NavyBlue]{hyperref}
\usepackage{color}
\usepackage{xcolor}
\usepackage{comment}

\DeclareMathAlphabet{\pazocal}{OMS}{zplm}{m}{n}
\usepackage{amsmath}
\usepackage{amssymb}
\usepackage[T1]{fontenc}

\usepackage{color}
\usepackage{makecell}

\begin{document}

\preprint{APS/123-QED}

\title{The Interplay of Thermal Melting and Pump Driven Melting of Charge Order:\\
A Two-Temperature Study of the Holstein Model}

\author{Debraj Bose}
\affiliation{Harish-Chandra Research Institute, Chhatnag Road,
Jhusi, Allahabad 211019}
\affiliation{Homi Bhabha National Institute, Training School
Complex, Anushakti Nagar, Mumbai 400 094, India}

\author{Sankha Subhra Bakshi}
\affiliation{Harish-Chandra Research Institute, Chhatnag Road, 
Jhusi, Allahabad 211019}
\affiliation{Homi Bhabha National Institute, Training School 
Complex, Anushakti Nagar, Mumbai 400 094, India}
\affiliation{Department of Physics, University of Virginia, 
Charlottesville, Virginia, 22904, USA}

\author{Pinaki Majumdar}
\affiliation{School of Arts and Sciences, Ahmedabad University, 
Navrangpura, Ahmedabad, India 380009}

\pacs{75.47.Lx}
\date{\today}

\begin{abstract}
Charge order driven by electron-phonon coupling is well understood at equilibrium 
but pump-probe experiments raise a new question: how does this order melt and 
recover after strong photoexcitation?  A pump pulse promotes carriers across the 
charge-order gap and creates a nonequilibrium high-energy electronic population.
In a closed system the subsequent dynamics is constrained by energy conservation.
In an `open system' - where the system is coupled to a thermal bath at some 
temperature $T_{\rm bath}$ -  there are new fluctuation and dissipation processes at
play. One can attempt a computational scheme that incorporates coupling of 
electrons to a laser pump, the coupling of system phonons to a thermal bath, 
and the Holstein interaction that couples electrons and phonons. We attempt
an approximation where the pump induced electronic excitations are  modeled 
by a slowly time varying `electron temperature', $T_{\rm el}(t)$, indicative of
a quasi-equilibrium electronic state. We solve the problem for different 
combinations of $T_{\rm el}$ and $T_{\rm bath}$, probing the order parameter dynamics, 
the static properties and excitations in the long time `quasi steady state', 
and establish a `phase diagram' in terms of bath temperature and 
electron temperature.
\end{abstract}

\maketitle

\section{Introduction}
\label{sec:introduction}

Pump-probe experiments provide a direct route to nonequilibrium dynamics in 
correlated electron systems \cite{Bloch,Beaure,Carpene08,Beaud,Fiebig,
Matsuzaki,cervantes2019,Iwai,
Okamoto,cho2015,Ichikawa,Liu,Budden,Fausti,Zhang20,Makler21,Schmitt11,
Wolf14,Han2015,Hellmann10}. 
 A short electromagnetic pulse deposits energy primarily in the electronic 
sector, and the subsequent evolution can be monitored through time-resolved 
photoemission, optical conductivity, electron diffraction, or x-ray scattering \cite{Carpene09,Sobota21,Eich14,Gedik17,Devereaux16,Mitrano20,Zong19}.   
The pumped state is not equivalent to an equilibrium state at some
higher temperature and the electronic distribution, lattice degrees of freedom, 
and order parameter can evolve on distinct time scales.  This separation underlies 
transient loss of order \cite{Carpene08,Beaud,Iwai,Matsuzaki,cervantes2019,
Fiebig,Okamoto,cho2015}, 
photoinduced or hidden phases \cite{Fausti,Ichikawa,Liu,Budden}, and large 
changes in transport and optical response \cite{Iwai,Okamoto}.

Charge-density-wave and charge-ordered electron-phonon systems are especially 
useful for studying this physics.  Strong electron-phonon coupling can bind 
carriers to local lattice distortions and form polarons \cite{frohlich,
latt-pol,Setvin21}.  At commensurate filling, these polarons can order 
into a charge-ordered state with a periodic modulation of both the electronic 
density and the lattice displacement \cite{adams,Batrouni21,chen2016}.  
Pump-probe measurements on such systems often show a rapid suppression of 
charge order followed by slower recovery.  Understanding this recovery 
requires a description that retains both the local amplitude dynamics of 
the lattice distortion and the real-space growth of ordered domains.

The spinless Holstein model is a minimal setting for this problem, but its 
nonequilibrium dynamics is challenging even without a bath.  After a pump 
or quench in a closed system, the total energy is conserved and the dynamics 
cannot be reduced to relaxation toward an externally imposed temperature.  
Previous studies of Holstein and related charge-order models have shown 
that the nonequilibrium electronic population strongly affects the lattice 
response: it can suppress the charge-order amplitude, seed spatial patterns, 
and slow the return to long-range order 
\cite{bakshi2024chargeorder,yang2024pattern,yang2025photoinduced,
yang2026pseudospin,jang2026suppressed}.  Recent machine-learning 
force approaches have also emphasized that the long-time dynamics 
of charge-density-wave order is limited by the need to follow 
large-scale domain growth while retaining the electronic contribution 
to the lattice forces \cite{cheng2023machine,yang2024enhanced,
bakshi2026mlcdw}.  The dynamics is therefore controlled not only by 
the absorbed energy, but also by how this energy is distributed among 
excited carriers, lattice distortions, and spatial textures.

The open-system problem, where the system is coupled to a 
thermal bath,  is harder.  In a material, both electrons 
and phonons exchange energy with environmental 
degrees of freedom. Imagine a classical bath coupled to the
system phonons. In that case one can either (i) retain the
environment as part of an enlarged `system' and do Hamiltonian
evolution, or (ii) trace out the environment and bring in a
stochastic force and dissipation acting on the system phonons.
Neither of these is easy for large lattices that are
needed to probe spatiotemporal dynamics in the driven system.

We use an approximate version of (ii) above to study the 
simultaneous effect of an external pump and a thermal bath.
This is best motivated at strong electron-phonon 
coupling where the gap is large
compared to the phonon frequency or 
the bath temperature. In this case
 relaxation across the gap is bottlenecked.  
The lattice can then evolve for a long time in the presence of a 
long-lived excited electronic population.  This motivates a 
two-temperature description in which the electronic sector is 
characterized by slowly varying temperature $T_{\rm el}(t)$, 
while 
the phonons remain coupled to a bath at $T_{\rm bath}$.
The $T_{\rm el}(t)$ of course has to be extracted from a
microscopic calculation like (i), we discuss this later.

With this approximation the problem looks more familiar
-  Langevin dynamics becomes 
a natural tool for the slow phonon sector.  
The reduction of a system coupled to many bath modes to friction 
and noise is a standard result of nonequilibrium statistical 
mechanics and open-system theory \cite{zwanzig2001nonequilibrium,
vankampen2007stochastic,
gardiner2009stochastic,kubo1966fluctuation,caldeira1983path,
bhattacharjee2016intermediate}.  In equilibrium, Langevin dynamics 
provides a controlled description of classical lattice fields 
coupled to a thermal bath and can be benchmarked against adiabatic 
Monte Carlo \cite{bhattacharyya2019langevin,bhattacharyya2020anharmonic}.  
It also gives access to real-space dynamics on lattices large enough to 
follow domain formation, coarsening, and slow recovery.  The central 
approximation in the present work is to extend this idea to a 
nonequilibrium two-temperature setting: the phonon coordinates 
obey Langevin dynamics at $T_{\rm bath}$, while the electronic 
contribution to the phonon potential is evaluated using 
$T_{\rm el}$.

In this paper we construct and analyze such a two-temperature Langevin 
scheme for the half-filled spinless Holstein model.  We first motivate 
the time-dependent electronic temperature from closed-system pump 
dynamics, where the excited electronic population can be fit to a 
Fermi distribution.  We then use a short-range phonon distortion 
dependent expression for the electronic energy, 
calibrated on  the two-site Holstein problem, to avoid 
repeated diagonalization of the full Hamiltonian 
during long Langevin simulations.  
The resulting model allows us to distinguish ordinary thermal 
melting from pump-induced melting.  The former is mainly a loss of 
spatial coherence between locally distorted regions, while the latter 
directly suppresses the local distortion amplitude and the associated 
electronic gap.

The paper is organized as follows.  In Sec.~\ref{sec:model_method} 
we introduce the microscopic electron-phonon-bath model, derive the 
stochastic mean-field equations, and explain the approximations that 
lead to the two-temperature Langevin dynamics.  In Sec.~\ref{sec:results} 
we present the results in two parts: (i)~transient suppression and 
recovery dynamics and (ii)~quasi-steady-state properties.  
In Sec.~\ref{sec:discussion} we 
discuss the limitations of the phenomenological electronic relaxation 
scheme, and Sec.~\ref{sec:conclusion} summarizes the main conclusions.

\section{Model and Method}
\label{sec:model_method}

Our goal is to reduce a microscopic driven electron-phonon-bath 
problem to a tractable dynamical model for charge-order recovery.  
The pump creates a nonequilibrium electronic population, while 
the lattice remains coupled to a thermal environment.  A complete 
treatment would require simultaneous evolution of the driven 
electrons, quantum phonons, the pump field, and the bath.  We 
instead start from this formulation and use two approximations: 
the pump-excited electrons are represented by an effective 
temperature, and the electronic free energy is approximated 
by a short-range phonon potential.

\subsection{Microscopic electron-phonon model}
\label{subsec:microscopic_model}

We consider spinless electrons locally coupled to dispersionless 
phonons on a two-dimensional square lattice.  Including the pump 
field and a phonon bath, the Hamiltonian is
\begin{align}
H &= H_{el}+H_{\rm ph}+H_{\rm el-ph}+H_{\rm bath}+H_{{\rm ph}\text{-}{\rm bath}} 
\nonumber\\
  &= \sum_{ij} t_{ij}(t)\hat c_i^\dagger \hat c_j
  + \sum_i \left({\hat p_i^2 \over 2M}+{K\hat x_i^2 \over 2}\right)
  - g\sum_i \hat n_i\hat x_i \nonumber\\
  &\quad + \sum_\alpha \left({\hat P_\alpha^2 \over 2M_\alpha}+{K_\alpha\hat 
X_\alpha^2 \over 2}\right)
  + \sum_{i\alpha} \kappa_{i\alpha}\hat x_i\hat X_\alpha .
\label{eq:full_hamiltonian}
\end{align}
Here $\hat x_i$ and $\hat p_i$ are the local phonon displacement and 
momentum, $M$ is the phonon mass, $K$ is the bare stiffness, and $g$ 
is the electron-phonon coupling. The bare phonon frequency is 
$\Omega_0=\sqrt{K/M}$. The bath variables $\hat X_\alpha$ and 
$\hat P_\alpha$ generate dissipation and thermal noise in the phonon sector 
through their coupling to the local phonon displacement $\hat x_i$ at site $i$, 
with coupling strength $\kappa_{i\alpha}$.
The pump is included through the Peierls substitution $t_{ij}(t)=
t_{ij}^0\exp[i\int_{\mathbf R_i}^{\mathbf R_j}\mathbf A(t)\cdot 
d\mathbf l]$, with $\mathbf E(t)=-\partial_t\mathbf A(t)$.  In 
the absence of the pump, $t_{ij}^0=-t_{\rm hop}$ for 
nearest-neighbour sites and zero otherwise.  We use $t_{\rm hop}$ 
as the unit of energy.

%%%%%%%%%%%%%%%%%%%%%%%%%%%%%%%%%%
\begin{figure*}[t]
\centerline{
\includegraphics[width=8.5cm,height=5cm]{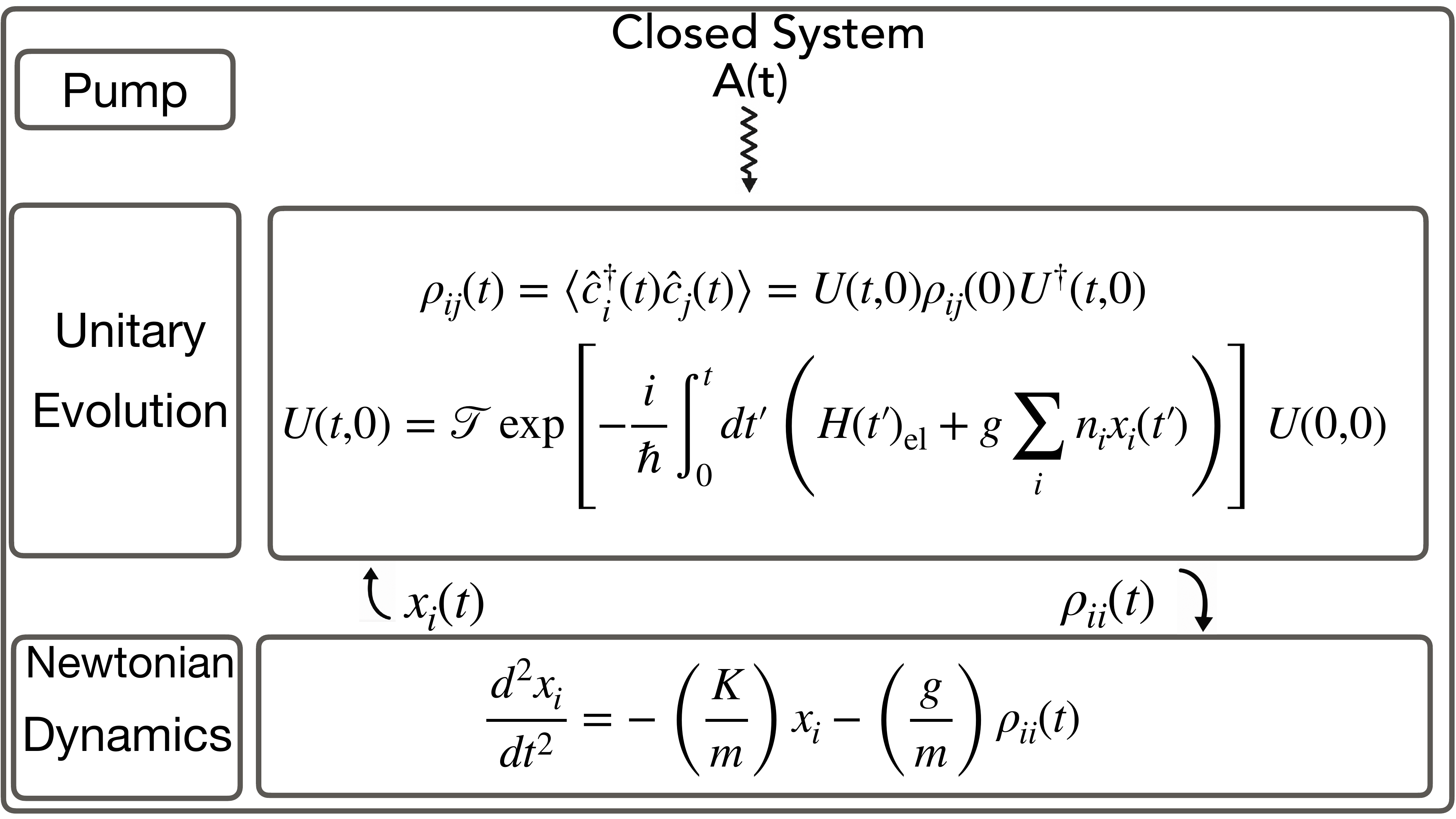}
\hspace{.3cm}
\includegraphics[width=8.5cm,height=5cm]{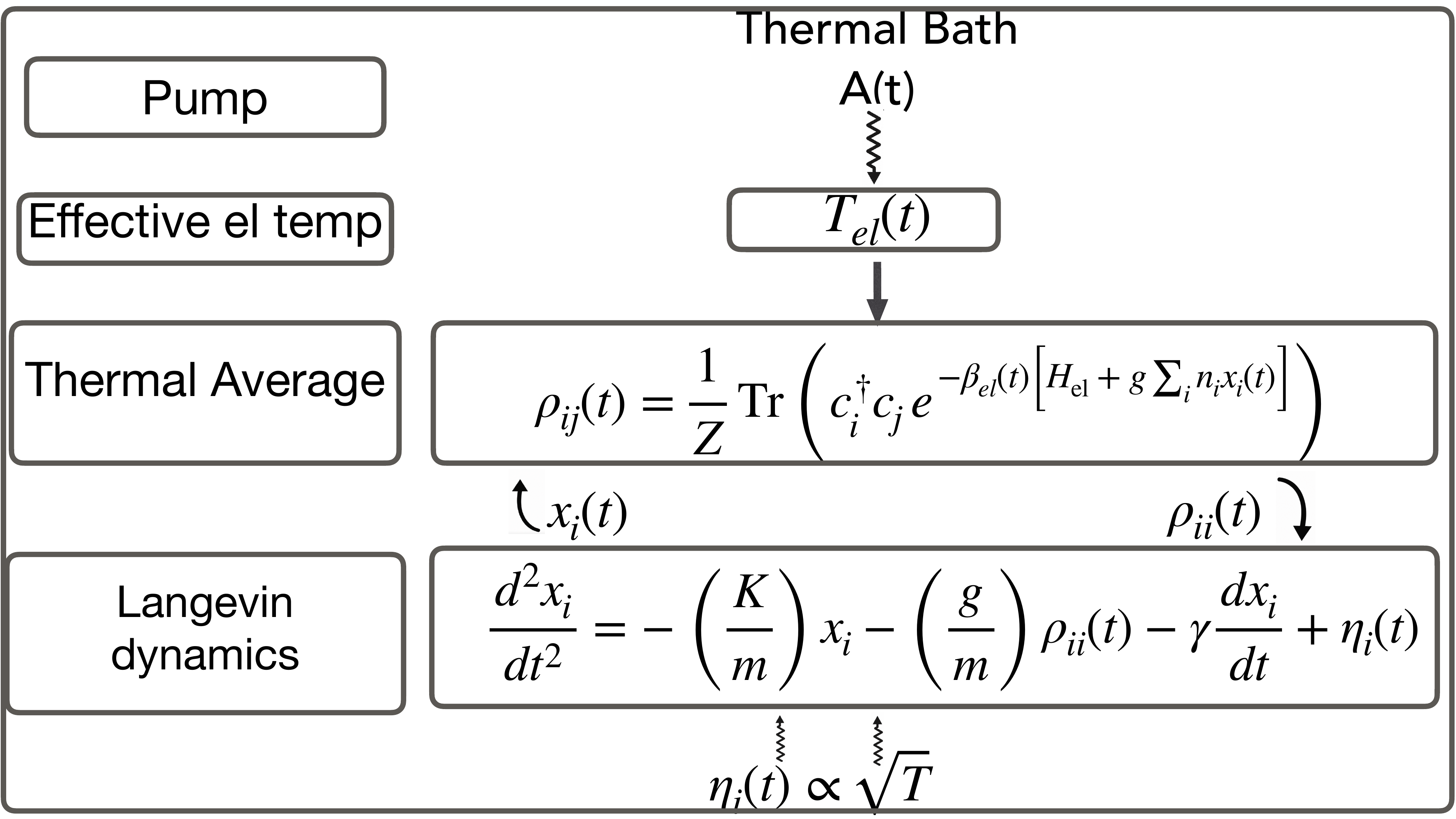}
}
\caption{Schematic of two methods. Left: the response of a closed system (no
thermal bath) to a pump pulse. The electrons `see' the pump pulse and the
phonon fluctuations, while the phonons see the fluctuating electron density. 
The system as a whole is energy conserving once the pump pulse passes. Right:
the `two temperature' version of an open system. Here the pump generates an
effective electron temperature $T_{\rm el}(t)$ and electronic properties are
computed as a {\it thermal average} at this temperature and in the background
of instantaneous phonons. The phonons in turn see the instantaneous electron
density. This scheme is valid provided $T_{\rm el}$ and the phonons vary on a
timescale much greater than electron hopping. 
}
\label{fig:flowchart}
\end{figure*}
%%%%%%%%%%%%%%%%%%%%%%%%%%%%%%%%%%

The Heisenberg equations generated by Eq.~\eqref{eq:full_hamiltonian} are
\begin{equation}
\begin{aligned}
\dot{\hat{x}}_i &= \frac{\hat{p}_i}{M}, \\[2pt]
\dot{\hat{p}}_i &= -K\hat{x}_i + g\hat{n}_i - \sum_\alpha 
\kappa_{i\alpha}\hat{X}_\alpha, \\[2pt]
\dot{\hat{X}}_\alpha &= \frac{\hat{P}_\alpha}{M_\alpha}, \\[2pt]
\dot{\hat{P}}_\alpha &= -K_\alpha \hat{X}_\alpha - \sum_i 
\kappa_{i\alpha}\hat{x}_i, \\[2pt]
\dot{\hat{\rho}}_{ij} &= i[\hat{\rho},t(t)]_{ij}
-i[\hat{\rho},g\hat{x}_i\delta_{ij}]_{ij}.
\end{aligned}
\label{eq:heisenberg_equations}
\end{equation}
Here $\hat\rho_{ij}=\hat c_i^\dagger\hat c_j$. These 
equations are exact at the operator level.

\subsection{Stochastic mean-field dynamics}
\label{subsec:smfd}

The exact equations do not close for expectation values.  For 
example, the equation for $\rho_{ij}=\langle \hat\rho_{ij}\rangle$ 
contains mixed electron-phonon correlators such as $\langle \hat 
x_i\hat\rho_{ij}\rangle$. Continuing their equations of motion 
generates the BBGKY hierarchy.  We close this hierarchy at the 
mean-field level by using $\langle \hat x_i\hat\rho_{ij}
\rangle\simeq x_i\rho_{ij}$, where $x_i=\langle\hat x_i\rangle$  
\cite{bakshi2024chargeorder}.
The lattice coordinates are then treated as classical dynamical 
variables, while the electronic density matrix evolves quantum 
mechanically in the instantaneous lattice background.

After this closure, the bath variables can be integrated out. 
Integrating out the harmonic bath gives a generalized Langevin 
equation with a memory kernel and fluctuating force; in the Ohmic, 
Markovian limit this reduces to local damping and Gaussian white 
noise satisfying the fluctuation-dissipation relation 
\cite{zwanzig2001nonequilibrium,vankampen2007stochastic,
gardiner2009stochastic,kubo1966fluctuation,caldeira1983path,
bhattacharjee2016intermediate}.  The resulting stochastic 
mean-field equations are
\begin{align}
M\ddot x_i &= -Kx_i+g\rho_{ii}-\gamma\dot x_i+\eta_i(t), \nonumber\\
\dot\rho_{ij} &= -i[h(\{x_i\},t),\rho]_{ij},\\
h_{ij}(\{x_i\},t) &=t_{ij}(t)-gx_i\delta_{ij} .
\label{eq:smfd}
\end{align}
The noise satisfies $\langle\eta_i(t)\rangle=0$ and $\langle\eta_i(t)
\eta_j(t')\rangle=2\gamma T_{\rm bath}\delta_{ij}\delta(t-t')$.  Thus 
the phonons exchange energy with a bath at $T_{\rm bath}$, while the 
electrons evolve in the time-dependent lattice and pump background.

Equation~\eqref{eq:smfd} is a microscopically motivated stochastic 
mean-field dynamics and is useful for studying the immediate effect 
of the pump.  It is, however, too expensive for the long-time 
simulations needed here.  The electronic density matrix contains 
${\cal O}(N^2)$ variables, and the adiabatic ratio $\Omega_0/t_{\rm hop}$ 
makes the phonon sector slow compared with the electronic hopping scale.  
A direct open-system calculation would therefore require a stable 
high-order scheme for coupled fast ODEs and slow stochastic 
differential equations over very long times.

In the closed-system limit, $\gamma=0$, the stochastic force is absent 
and Eq.~\eqref{eq:smfd} reduces to a coupled set of ordinary 
differential equations.  These equations can be solved accurately 
using, for example, a fourth-order Runge--Kutta algorithm, and 
they provide the key input for our reduced description: the 
electronic distribution produced by the pump.  For the open-system 
simulations, we therefore make two simplifications guided by the 
closed-system dynamics.  First, we replace the explicit post-pump 
electronic evolution by a time-dependent effective electronic 
temperature $T_{\rm el}(t)$.  Second, we use $T_{\rm el}$ to 
construct an effective phonon potential, so that the long-time 
recovery can be studied through classical Langevin dynamics
of the lattice alone.
We provide a schematic of the working of the closed system dynamics
and the two temperature open system dynamics (see below) in Fig.1

%%%%%%%%%%%%%%%%%%%%%%%%%%%%%%%%%%
\begin{figure*}[t]
\centerline{
\includegraphics[width=17cm,height=5cm]{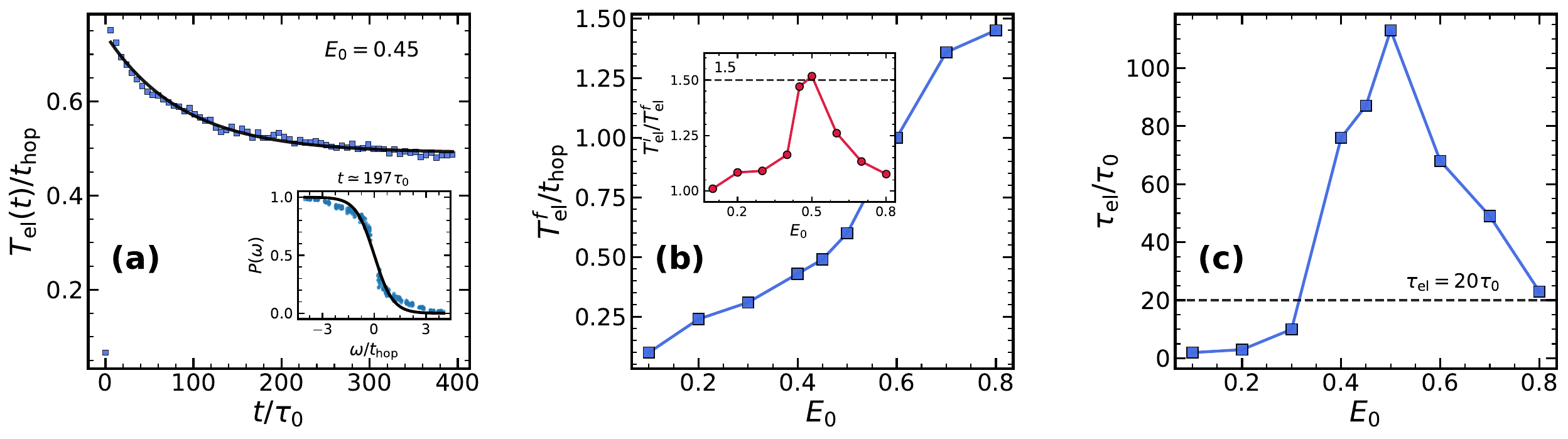}
}
\caption{Calibration of the electronic-temperature quench 
from closed-system pump dynamics.  (a) $T_{\rm el}(t)$ 
extracted from the instantaneous electronic population for 
$E_0=0.45$, with an exponential fit.  The inset shows a 
representative Fermi-function fit to $P(\omega)$.  (b) 
$T^f$ (and the ratio $T^i_{\rm el}/T^f_{\rm el}$ in the inset) as a function of 
the pump strength $E_0$. (c) Extracted relaxation time $\tau_{\rm el}$.}
\label{fig:Tel_calibration}
\end{figure*}
%%%%%%%%%%%%%%%%%%%%%%%%%%%%%%%%%%

\subsection{Approximations}
\label{subsec:approximations}

\subsubsection{Electronic-temperature quench}
\label{subsubsec:Tel_quench}

The pump primarily excites the electronic sector and creates a 
nonequilibrium population of high-energy electron-hole excitations. 
 In a full microscopic treatment, one would follow the pulse-driven 
electronic dynamics, energy redistribution within the electronic 
sector, and energy transfer to the lattice and bath.  Here we 
retain only the net effect of this process by introducing a 
time-dependent electronic temperature $T_{\rm el}(t)$, while the 
phonons remain coupled to a bath at $T_{\rm bath}$.  The post-pump 
state is therefore described by a two-temperature condition, 
$T_{\rm el}(t)\ne T_{\rm bath}$.  The electronic temperature 
controls the electronic part of the phonon potential, while 
$T_{\rm bath}$ controls dissipation and thermal noise.

In principle one can calibrate the form of $T_{\rm el}(t)$ 
using the closed-system limit~\cite{bakshi2024chargeorder}.  
Setting $\gamma=0$, we apply an oscillating electric field with carrier
frequency $\omega_{\rm pump}=t_{\rm hop}$ and amplitude $E_0$. The pulse has
a Gaussian envelope characterized by $\Omega_{\rm pump}=3t_{\rm hop}$ and is
polarized along the $x+y$ direction, parallel to the system.
We extract the electronic population as a function of 
single-particle energy at each time.  This population is fit 
to a Fermi function, 
$$P(\omega,t)\simeq\{\exp[(\omega
-\mu(t))/T_{\rm el}(t)]+1\}^{-1},$$
 which defines the effective electronic temperature.  
Fig.\ref{fig:Tel_calibration}(a) 
shows a representative result, and the inset shows 
that the instantaneous population 
is well described by a Fermi form over the relevant energy range.

The extracted electronic temperature relaxes from 
a high initial value to a lower 
long-time value.  We parametrize this relaxation as
\begin{equation}
T_{\rm el}(t)=T_{\rm el}^f+\left(T_{\rm el}^i-
T_{\rm el}^f\right)e^{-t/\tau_{\rm el}} .
\label{eq:Tel_time}
\end{equation}
Here $T_{\rm el}^i$ is the post-pump value, $T_{\rm el}^f$ is the long-time 
value, and $\tau_{\rm el}$ is the electronic relaxation time.  Repeating the 
procedure for several pump amplitudes gives the calibration curves in 
Figs.\ref{fig:Tel_calibration}(b) and \ref{fig:Tel_calibration}(c).  
Thus $E_0$ fixes the effective temperature-quench parameters $T_{\rm el}^i$, 
$T_{\rm el}^f$, and $\tau_{\rm el}$.

In the open-system simulations below, we assume that the same functional 
form remains a useful phenomenological description.  To keep the parameter 
space small, we fix $T_{\rm el}^i/T_{\rm el}^f=1.5$ and $\tau_{\rm el}=
20\tau_0$, and use $T_{\rm el}^f$ as the main measure of pump strength.

\subsubsection{Effective short-range phonon model}
\label{subsubsec:effective_model}

The remaining task is to compute the force on the lattice distortions 
without diagonalizing the full electronic Hamiltonian at every Langevin 
step.  We use a short-range effective potential motivated by the 
two-site Holstein problem.  This approximation avoids the ${\cal O}(N^3)$ 
cost of repeated diagonalization.  More flexible force models, including 
machine-learning constructions~\cite{bakshi2026mlcdw}, could improve the 
quantitative accuracy, but the two-site form is sufficient for capturing 
the strong-coupling physics that is central to this work.

For one electron on two sites,
\begin{equation}
H_2^{\rm el}=-t_{\rm hop}\left(c_1^\dagger c_2+c_2^\dagger c_1\right)
-g(n_1x_1+n_2x_2).
\label{eq:two_site_hamiltonian}
\end{equation}
The two eigenvalues are $\lambda_\pm(x_1,x_2)=
-gx_+/2\pm\sqrt{g^2x_-^2/4+t_{\rm hop}^2}$, with $x_+=x_1+x_2$ 
and $x_-=x_1-x_2$.  At electronic temperature $T_{\rm el}$ the 
corresponding two-site electronic free energy is
\begin{equation}
F_{\rm el}(x_1,x_2;T_{\rm el})=-T_{\rm el}\ln
\left(e^{-\beta_{\rm el}\lambda_+}+e^{-\beta_{\rm el}\lambda_-}\right),
\label{eq:two_site_free_energy}
\end{equation}
where $\beta_{\rm el}=1/T_{\rm el}$.
We use this free energy as a nearest-neighbour interaction 
between lattice distortions.  The effective phonon 
Hamiltonian is
\begin{equation}
H_{\rm ph}^{\rm eff}=\sum_i {p_i^2\over2M}+{K\over2}\sum_i 
x_i^2+\sum_{\langle ij\rangle}F_{\rm el}(x_i,x_j;T_{\rm el}).
\label{eq:effective_phonon_hamiltonian}
\end{equation}

This form retains the bare local stiffness and includes the 
short-range electronic tendency toward alternating distortions.  
Increasing $T_{\rm el}$ weakens the effective ordering tendency 
and thereby represents the effect of pump-excited carriers.

The final lattice dynamics is then
\begin{equation}
M\ddot x_i=-Kx_i-{\partial\over\partial x_i}
\sum_{j\in NN(i)}F_{\rm el}(x_i,x_j;T_{\rm el}(t))
-\gamma\dot x_i+\eta_i(t),
\label{eq:final_langevin}
\end{equation}
with $\langle\eta_i(t)\rangle=0$ and $\langle
\eta_i(t)\eta_j(t')\rangle=2\gamma T_{\rm bath}
\delta_{ij}\delta(t-t')$.  In the simulations we 
set $t_{\rm hop}=1$, $g=2$, $K=1$, and $\Omega_0
=0.2t_{\rm hop}$, giving $E_p=g^2/(2K)$ and $\tau_0=2\pi/\Omega_0$.  
Unless stated otherwise, the damping is $\gamma=0.05$, the lattice 
size is $L=30$--$50$, and observables are averaged over 
independent noise realizations.

Before applying the two-temperature scheme, we check the equilibrium 
limit.  When $T_{\rm el}=T_{\rm bath}=T$, the Langevin distribution 
generated by Eq.~\eqref{eq:final_langevin} should reproduce the 
thermal properties of the adiabatic Holstein model, at least 
qualitatively.  The charge-order wave vector is ${\bf Q}=(\pi,
\pi)$, and we define $S_{\bf Q}=|N^{-1}\sum_i x_i e^{i{\bf Q}
\cdot{\bf R}_i}|^2$.

In the Appendix we compare $S_{\bf Q}$ from the 
two site potential with the corresponding Holstein 
Monte Carlo result.

\section{Results}
\label{sec:results}

We now present the results of the two-temperature Langevin dynamics. 
 We first discuss transient pump dynamics, including the order 
parameter, gap, recovery timescales, and real-space domain 
evolution.  Then, we analyze quasi-steady states in the 
$T_{\rm el}$--$T_{\rm bath}$ plane.

\subsection{Transient dynamics}
\label{subsec:transient}

We next consider the pump protocol.  The electronic temperature 
relaxes from $T_{\rm el}^i$ to $T_{\rm el}^f$ according to 
Eq.~\eqref{eq:Tel_time}, while the phonons remain coupled to a 
bath at $T_{\rm bath}$.  We use $T_{\rm el}^f/t_{\rm hop}=0.3$,
 $0.5$, and $0.7$ as representative weak, intermediate, and 
strong pumps.  The transient simulations are performed on 
$50\times50$ lattices unless stated otherwise.  In comparing
 different parameters, $S_{\bf Q}(t)=|N^{-1}\sum_i 
x_i(t)e^{i{\bf Q}\cdot{\bf R}_i}|^2$ is normalized by its 
low-temperature equilibrium value.

%%%%%%%%%%%%%%%%%%%%%%%%%%
\begin{figure}[b]
\centerline{
\includegraphics[width=8.5cm,height=5.2cm]{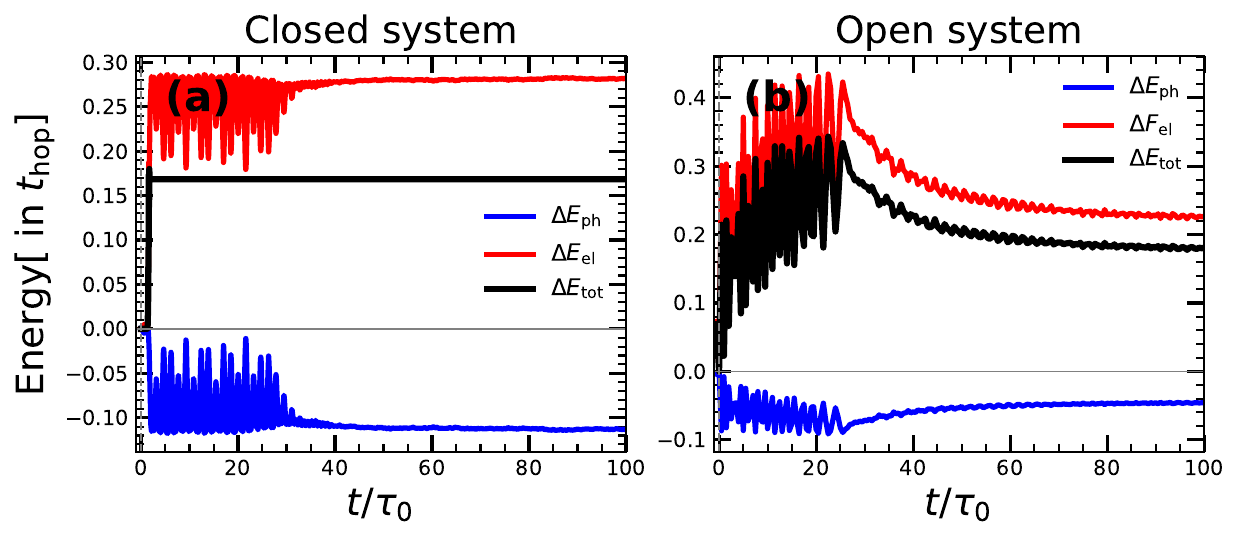}
}
\caption{
Energy dynamics after the pump for closed and open systems at
$E_0=0.47$.  The open system has $T_{\rm el}^f=0.5t_{\rm hop}$ and
$T_{\rm bath}=0.05t_{\rm hop}$, and relaxes to a two-temperature
quasi-steady state rather than to equilibrium at $T_{\rm bath}$.
}
\label{fig:energy_dynamics}
\end{figure}
%%%%%%%%%%%%%%%%%%%%%%%%%%
\begin{figure}[t]
\centerline{
\includegraphics[width=8.5cm,height=4.8cm]{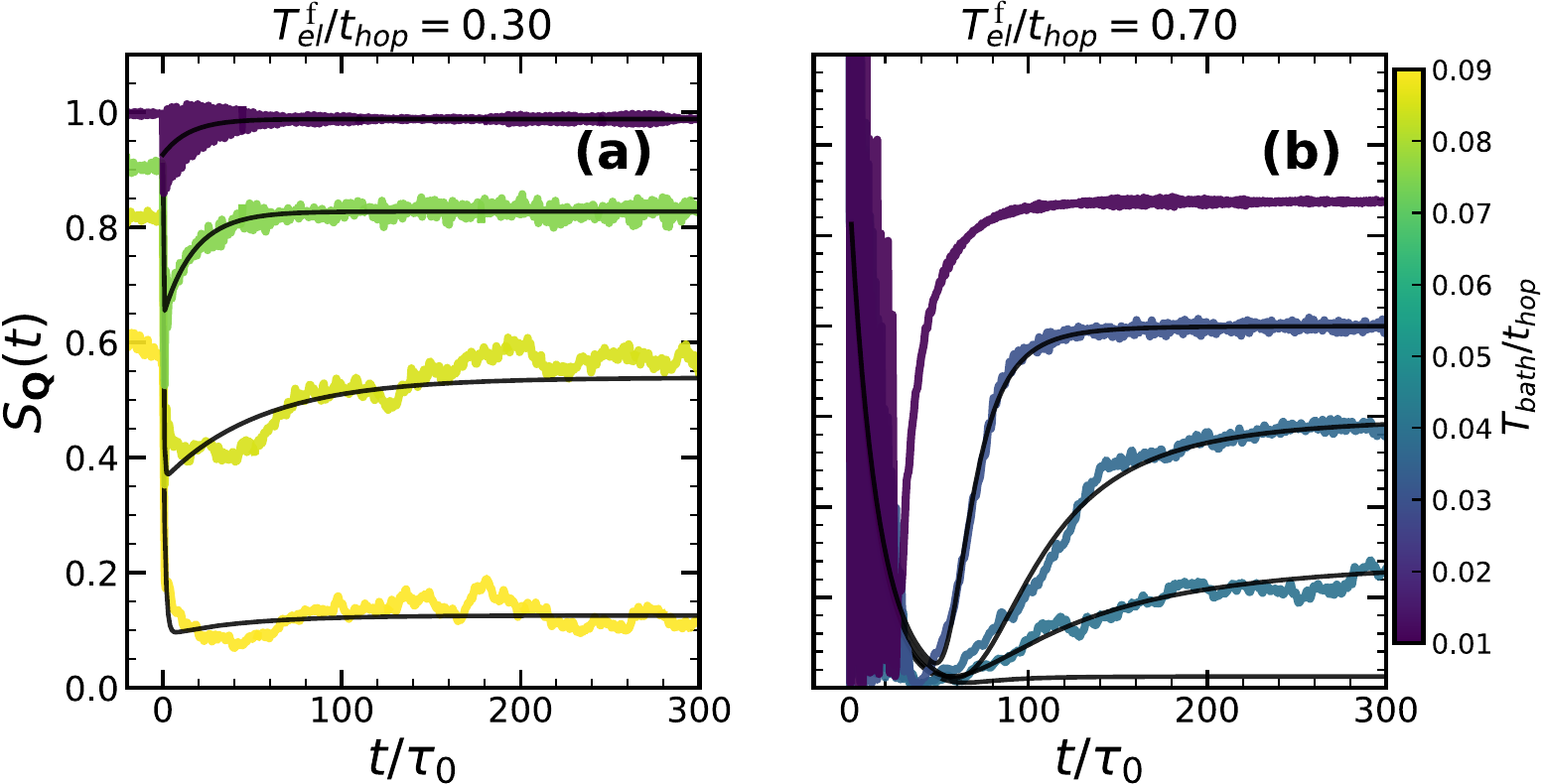}
}
\caption{Transient loss and recovery of charge order.  Weak
pumping, $T_{\rm el}^f/t_{\rm hop}=0.3$, suppresses $S_{\bf Q}$
without fully destroying it, while strong pumping,
$T_{\rm el}^f/t_{\rm hop}=0.7$, can nearly extinguish the order.
Left panels show individual trajectories and right panels
show noise-averaged results.}
\label{fig:transient_recovery}
\end{figure}
%%%%%%%%%%%%%%%%%%%%%%%%%%
%%%%%%%%%%%%%%%%%%%%%%%%%%
\begin{figure}[b]
\centerline{
\includegraphics[width=6.5cm,height=4.8cm]{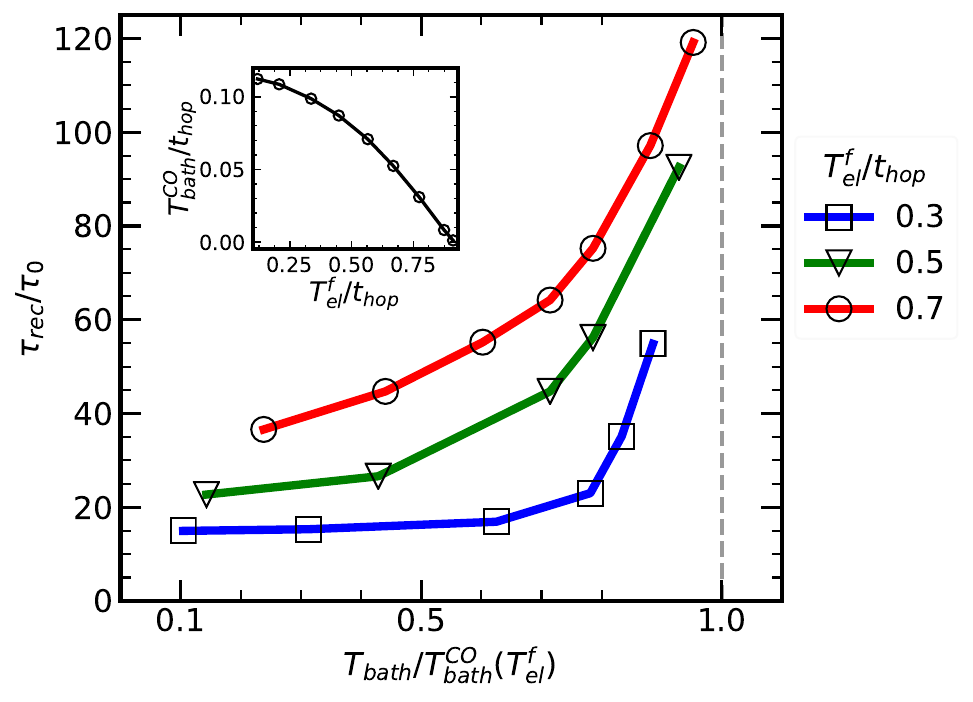}
}
\caption{Recovery time $\tau_{\rm rec}$ as a function of the bath temperature, normalized by the corresponding charge-ordering transition temperature $T^{\rm CO}_{\rm bath}(T^f_{\rm el})$, for different final electronic temperatures $T^f_{\rm el}$. The inset shows $T^{\rm CO}_{\rm bath}$ as a function of $T^f_{\rm el}$.}

\label{fig:gap_tau_rec}
\end{figure}
%%%%%%%%%%%%%%%%%%%%%%%
\subsubsection{Time dependence of the energy}
\label{subsubsec:energy_dynamics}

Before discussing the order-parameter recovery, we first examine the
energy flow after the pump.  This is useful because the closed and open
systems differ most directly in how the absorbed energy is redistributed
and removed.  In the closed system, the pump injects energy into the
electronic sector.  After the pulse has passed, the total energy is
conserved, although the electronic and phonon parts continue to exchange
energy through the Holstein coupling.  Thus the closed-system dynamics
describes redistribution of a fixed amount of absorbed energy.

The open-system dynamics is different.  The phonons are coupled to a bath
at $T_{\rm bath}$ and therefore lose energy through the dissipative part of
the Langevin dynamics, while the stochastic force maintains thermal
fluctuations at the bath temperature.  At the same time, the electronic
sector is not allowed to cool self-consistently to the bath.  Instead, its
effect on the phonon potential is controlled by the imposed electronic
temperature $T_{\rm el}(t)$, which relaxes to the long-time value
$T_{\rm el}^f$.  The final state is therefore a two-temperature
quasi-steady state, not an equilibrium state at $T_{\rm bath}$.

%%%%%%%%%%%%%%%%%%%%%%%%%%%%
\begin{figure*}[t]
\centerline{
\includegraphics[width=17cm,height=10cm]{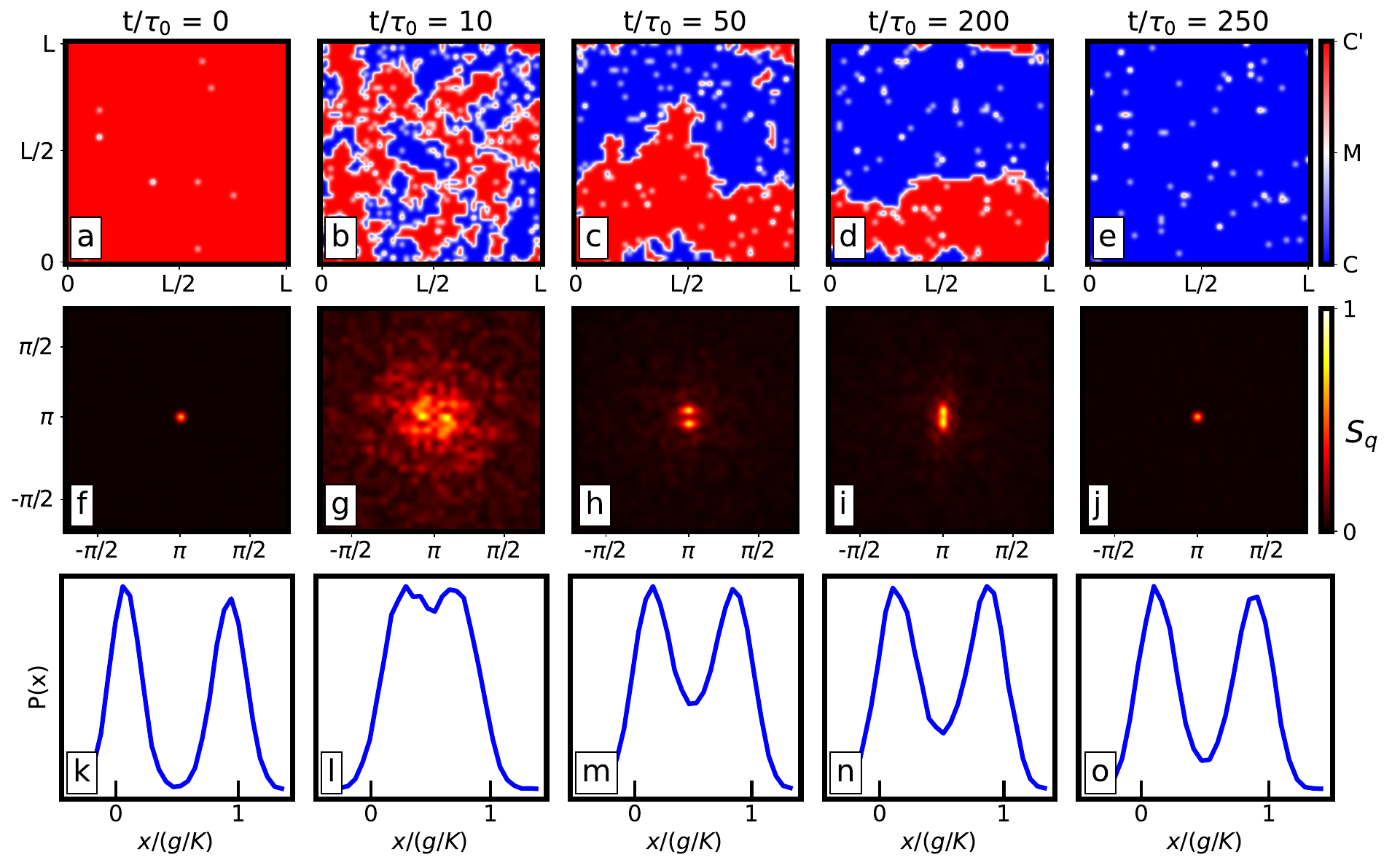}
}
\caption{Real-space recovery after an intermediate pump,
$T_{\rm el}^f/t_{\rm hop}=0.5$, at $T_{\rm bath}/t_{\rm hop}
=0.05$.  The pump creates oppositely phased charge-order
domains and locally melted regions; subsequent coarsening
restores a dominant checkerboard sector.  Momentum-space
weight sharpens at ${\bf Q}$ only after local bimodality
in $P(x)$ has largely recovered.}
\label{fig:domain_dynamics}
\end{figure*}
%%%%%%%%%%%%%%%%%%%%%%%%%%%%%%%%%%%%%%%%%

Fig.~\ref{fig:energy_dynamics} compares the time dependence of the energy 
per site relative to the corresponding low-temperature equilibrium value,
\(\Delta E(t)=E(t)-E_{\rm eq}\), in the closed and open cases for the 
same pump strength. In the closed
case, the total energy remains constant after the pump, as expected for
Hamiltonian dynamics.  In the open case, the energy decreases after the
pump because the lattice is coupled to the bath.  However, the system does
not relax back to the low-temperature equilibrium energy set only by
$T_{\rm bath}$.  Instead, it approaches a higher quasi-steady value
controlled by the residual electronic temperature $T_{\rm el}^f$.  This
energy relaxation illustrates the central assumption of the present
approach: the bath cools the phonon sector, while a long-lived hot
electronic population continues to weaken the charge-ordering tendency.

\subsubsection{Order-parameter dynamics}
\label{subsubsec:order_parameter}

Fig.\ref{fig:transient_recovery} shows the time evolution of 
the charge-order structure factor for weak and strong pumps.  For
 $T_{\rm el}^f/t_{\rm hop}=0.3$, the order parameter is suppressed
 after the pump but remains finite.  The system retains memory 
of the original checkerboard pattern and recovers smoothly after 
noise averaging.  For $T_{\rm el}^f/t_{\rm hop}=0.7$, the 
suppression is much stronger: $S_{\bf Q}$ can collapse close 
to zero, and the recovery becomes slower and more trajectory 
dependent.  The strong-pump case therefore involves 
reconstruction of charge order from a substantially disordered 
state, not merely repair of a weakly damaged pattern.

\subsubsection{Recovery timescales}
\label{subsubsec:timescales}

We extract a characteristic recovery time by fitting the 
noise-averaged trajectories.  For weak pumping, where the 
order parameter remains finite, we use $S_{\bf Q}(t)=
A_de^{-t/\tau_{\rm decay}}+S_d+S_r(1-e^{-t/\tau_{\rm rec}})$. 
 Here $S_d$ is the damaged value of the order parameter, 
$S_r$ is the recovered component, and $\tau_{\rm rec}$ is
 the recovery time.  For strong pumping, where the order
 parameter can approach zero, the delayed growth is better 
fit by $S_{\bf Q}(t)=A_de^{-t/\tau_{\rm decay}}+
S_r\exp[-(\tau_{\rm rec}/t)^\alpha]$, with $\alpha\sim3\pm1$. 
 We use this exponent only as a fitting parameter, not as 
a claim of universal scaling.

The extracted timescales are shown in Fig.\ref{fig:gap_tau_rec}. 
 At low bath temperature, $\tau_{\rm rec}$ increases with pump 
strength because larger $T_{\rm el}^f$ produces a smaller residual 
distortion amplitude and a more disrupted domain configuration.  
As $T_{\rm bath}$ approaches the pump-dependent charge-ordering
 temperature $T_{\rm bath}^{\rm CO}(T_{\rm el})$, the recovery 
time grows rapidly.  This indicates critical slowing down near 
the nonequilibrium ordering boundary.  The boundary itself 
shifts to lower $T_{\rm bath}$ with increasing $T_{\rm el}$, 
showing that a hot electronic population reduces the bath 
temperature range over which long-range charge order can survive.

\subsubsection{Real-space dynamics}
\label{subsubsec:real_space}

The recovery of $S_{\bf Q}$ involves both local amplitude restoration 
and domain growth.  To visualize this process, we study an intermediate
 pump, $T_{\rm el}^f/t_{\rm hop}=0.5$, at $T_{\rm bath}/t_{\rm hop}=0.05$. 
 The checkerboard state has two symmetry-related configurations, $C$ and 
$C'$, differing by a sublattice phase shift.  We distinguish them using 
the local variable $\xi_i=(x_i/x_0-1)e^{i{\bf Q}\cdot{\bf R}_i}$, with 
$x_0=g/(2K)$.  In an ideal charge-ordered state, $\xi_i$ has opposite 
signs in the two domains; regions with strongly suppressed distortions 
correspond to locally melted patches.

Fig.\ref{fig:domain_dynamics} 
shows real-space snapshots, the corresponding momentum-space structure, 
and the local distortion distribution.  The initial state is nearly 
uniform.  Shortly after the pump, the original order is disrupted and 
domains of both $C$ and $C'$ appear, separated by domain walls and 
locally melted regions.  The Fourier peak at ${\bf Q}$ is replaced 
by broad weight associated with finite domains.  At later times the
 domains coarsen, one charge-order sector dominates, and the sharp 
${\bf Q}$ peak is restored.

The bottom row of Fig.\ref{fig:domain_dynamics} shows that the 
local distortion distribution recovers earlier than the global 
order.  Immediately after the pump, $P(x)$ broadens and the two 
peaks strongly overlap.  The bimodality returns before the system 
is globally ordered, confirming that local polaronic distortions 
recover before long-range phase coherence.

%%%%%%%%%%%%%%%%%%%%%%%%%%%%%%%%%%
\begin{figure}[t]
\centerline{
\includegraphics[width=5.5cm,height=4.8cm]{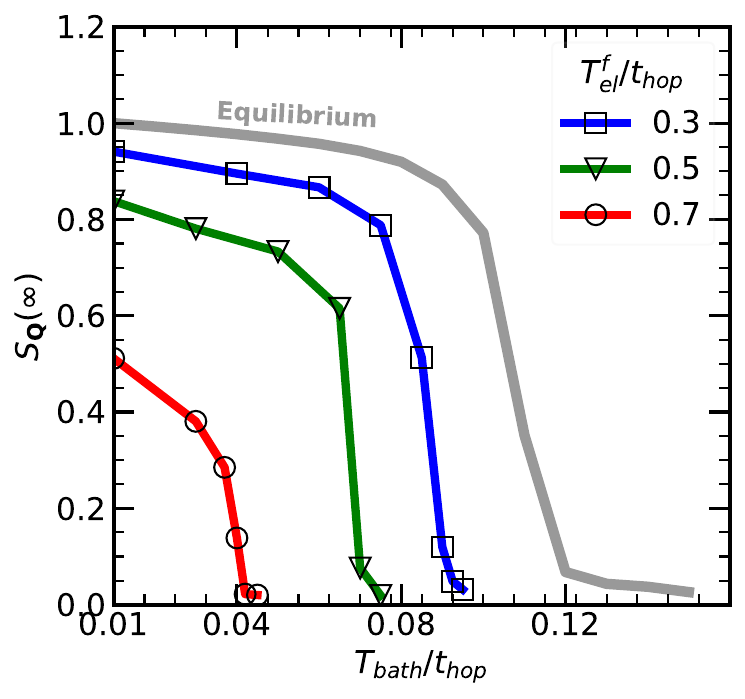}
}
\caption{Steady-state charge order. Long-time
structure factor $S_{\bf Q}$ versus bath
temperature for different $T_{\rm el}^f$}
\label{fig:steady_SQ_Px}
\end{figure}
%%%%%%%%%%%%%%%%%%%%%%%%%%%%%%%%%

\subsection{Steady state results}
After the transient, the system reaches a long-lived
quasi-steady state specified by the final electronic 
temperature $T^f_{\rm el}$ and the bath temperature $T_{\rm bath}$. 
This state is not in thermal equilibrium unless  $T^f_{\rm el}=T_{\rm bath}$
but it is stationary on the simulation time scale because the 
electronic population is held fixed through $T^f_{\rm el}$. 
The two temperatures affect charge order in different ways.
The bath temperature controls stochastic lattice fluctuations
and therefore mainly disorders the phase/domain structure.  The
electronic temperature enters through the effective phonon potential
and directly weakens the local distortion amplitude.

\subsubsection{Steady-state statics}
\label{subsubsec:steady_statics}

Fig.~\ref{fig:steady_SQ_Px} shows the long-time value of
$S_{\bf Q}$ as a function of $T_{\rm bath}$ for several
$T_{\rm el}^f$.  The gray curve is the equilibrium result, with a
thermal transition near $T_{\rm bath}/t_{\rm hop}\simeq0.12$.
For fixed nonzero $T_{\rm el}^f$, increasing $T_{\rm bath}$ still
destroys long-range order, but the transition shifts to lower bath
temperature.  Thus a hot electronic population reduces the range of
bath temperatures over which charge order can survive.  In addition,
even at $T_{\rm bath}=0$, $S_{\bf Q}$ decreases with increasing
$T_{\rm el}^f$.  This reduction cannot come from thermal phonon
fluctuations; it reflects the direct suppression of the local
charge-order distortion by the excited electronic population.

These results show that the loss of charge order is controlled by two
distinct mechanisms: bath-induced loss of spatial coherence and
electron-induced reduction of the local distortion amplitude.  We now
organize these regimes in the full $T_{\rm el}^f$--$T_{\rm bath}$
plane.
%%%%%%%%%%%%%%%%%%
\begin{figure}[b]
\centerline{
\includegraphics[width=5.5cm,height=4.8cm]{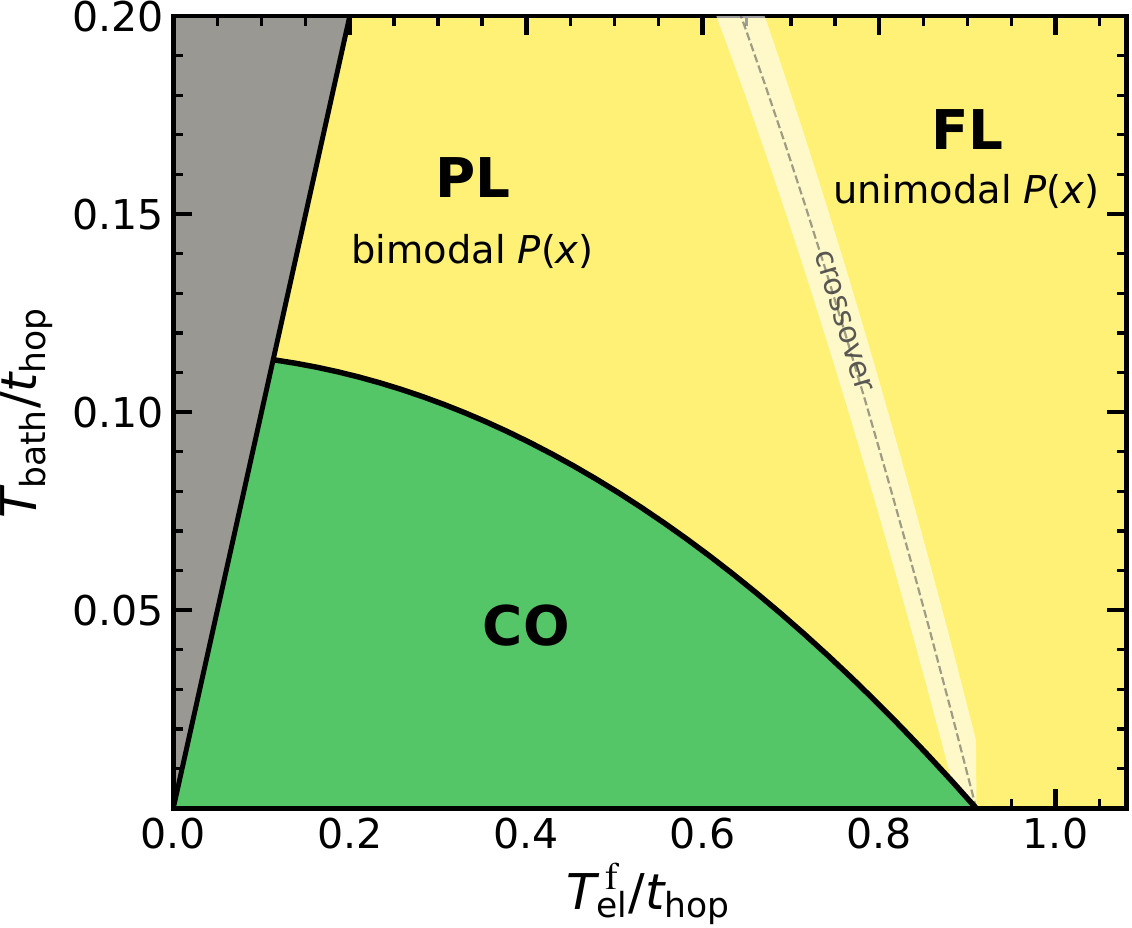}
}
\caption{
Quasi-steady-state phase diagram in the
$T_{\rm el}^f$--$T_{\rm bath}$ plane.  The low-temperature CO phase is
separated from the disordered regime by the charge-ordering boundary.
The shaded band marks the crossover from a polaron liquid with bimodal
$P(x)$ to a more homogeneous Fermi-liquid-like regime. The gray region denotes the part of parameter space not accessed by the pump protocol considered here.
}
\label{fig:phase_diagram}
\end{figure}
%%%%%%%%%%%%%%%%%%%%%%%%%%%
\subsubsection{Phase diagram}
\label{subsubsec:phase_diagram}

Fig.~\ref{fig:phase_diagram} summarizes the quasi-steady-state
behavior.  The low-temperature region is charge ordered.  Increasing
$T_{\rm bath}$ at small $T_{\rm el}^f$ destroys long-range order
through thermal lattice fluctuations, but local distortions can remain
finite.  We identify this disordered but locally distorted regime as a
polaron liquid.  Increasing $T_{\rm el}^f$ at low $T_{\rm bath}$
instead suppresses the distortion amplitude itself.  At sufficiently
large $T_{\rm el}^f$, the system crosses over toward a weakly
distorted, Fermi-liquid-like regime.

%%%%%%%%%%%%%%%%%%%%%%%%%%%%
\begin{figure*}[t]
\centerline{
\includegraphics[width=16cm,height=4.0cm]{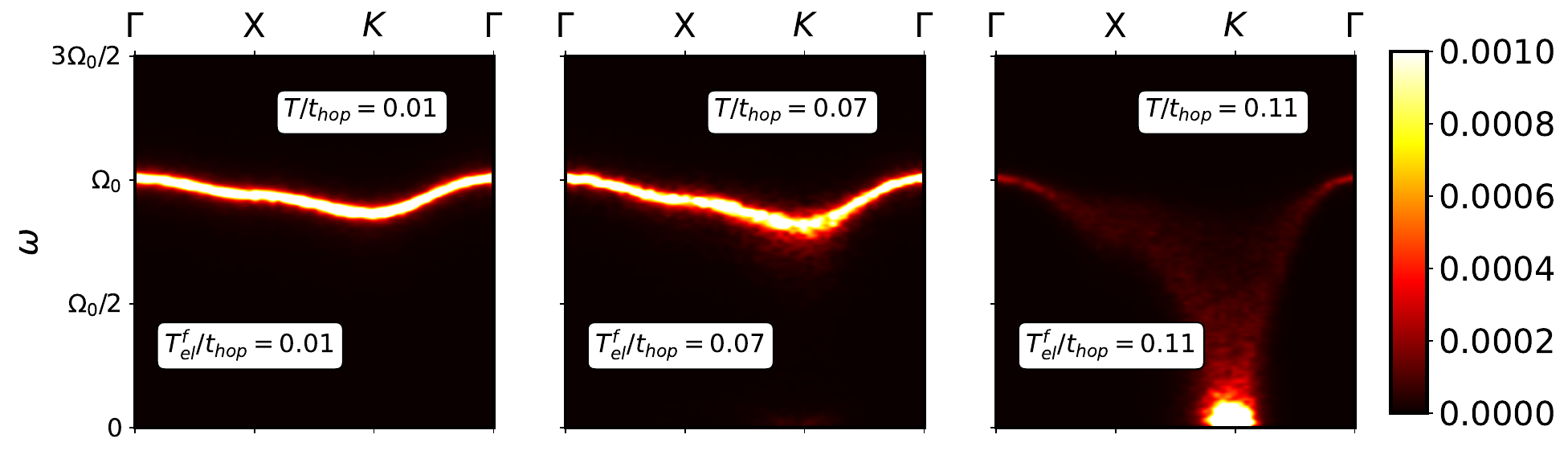}
}
\centerline{
\includegraphics[width=16cm,height=4.0cm]{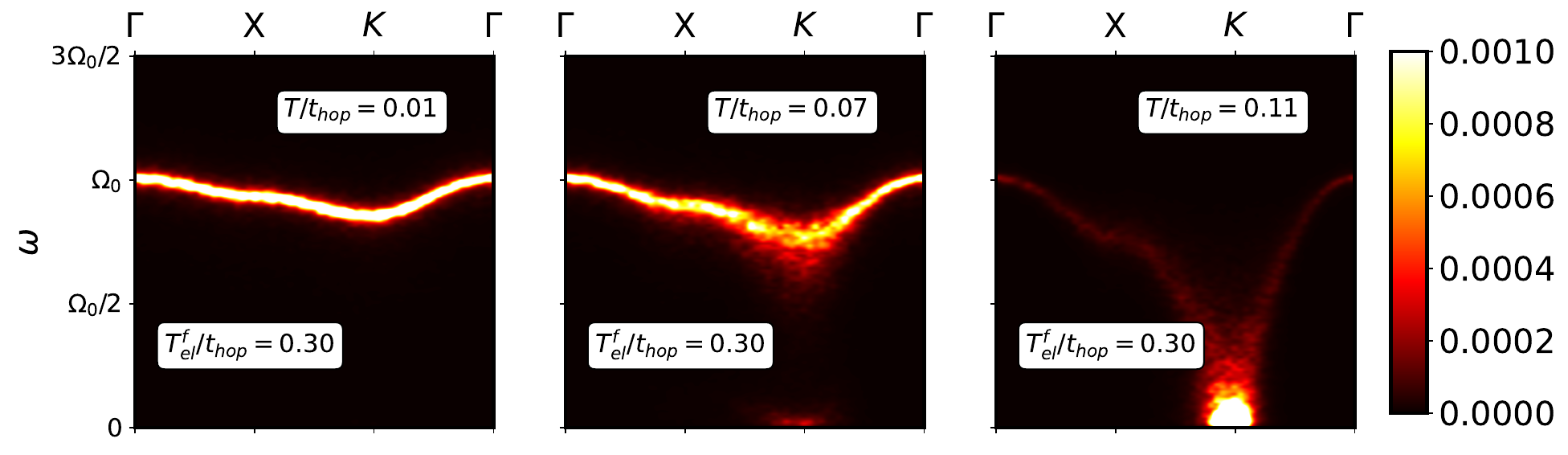}
}
\centerline{
\includegraphics[width=16cm,height=4.0cm]{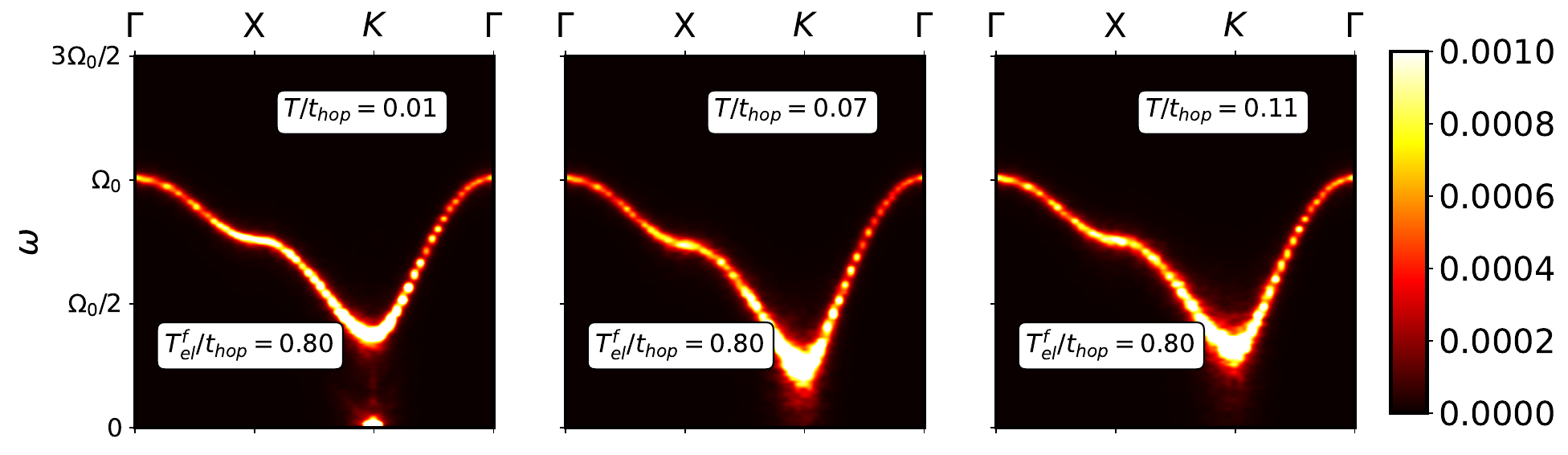}
}
\caption{Phonon dynamical structure factor along
$\Gamma(0,0)$--$X(\pi,0)$--$K(\pi,\pi)$--$\Gamma(0,0)$.
Rows show equilibrium, weak pump, and strong pump
conditions.  The charge-order mode softens and
broadens near the ordering boundary $T_{\rm CO}^{\rm bath}(T_{\rm el}^f)$ with stronger effects at larger $T^f_{\rm el}$}

\label{fig:phonon_dsf}
\end{figure*}
%%%%%%%%%%%%%%%%%%%%

The boundary of the CO phase is obtained from the loss of
$S_{\bf Q}$ in the long-time state.  The shaded band marks the
crossover where the local distortion distribution changes from
bimodal to nearly unimodal.  Thus the phase diagram separates two
different routes out of the ordered state: a thermal route into a
polaronic liquid and an electronic route toward a more homogeneous
metallic state.

The diagonal line denotes equilibrium, $T_{\rm el}^f=T_{\rm bath}$.
The gray region with $T_{\rm bath}>T_{\rm el}^f$ is outside the pump
protocol considered here, since after photoexcitation the electronic
sector is expected to remain hotter than the bath on the intermediate
time scale described by the model.

%%%%%%%%%%%%%%%%%%%%%%%%%%%
\begin{figure*}[t]
\centerline{
\includegraphics[width=17cm,height=4.5cm]{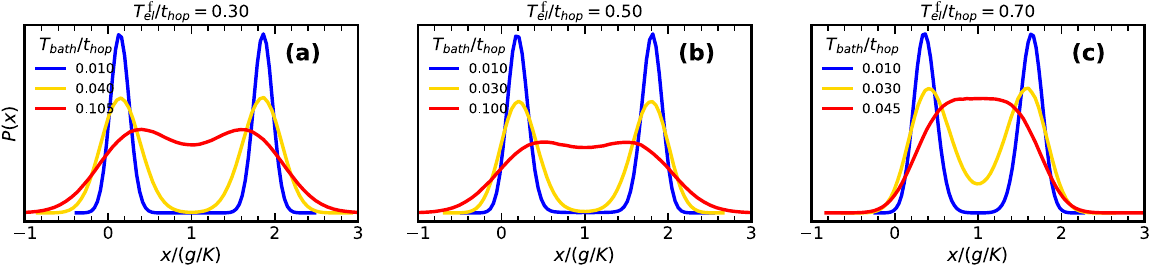}
}
\centerline{
\includegraphics[width=17cm,height=4.5cm]{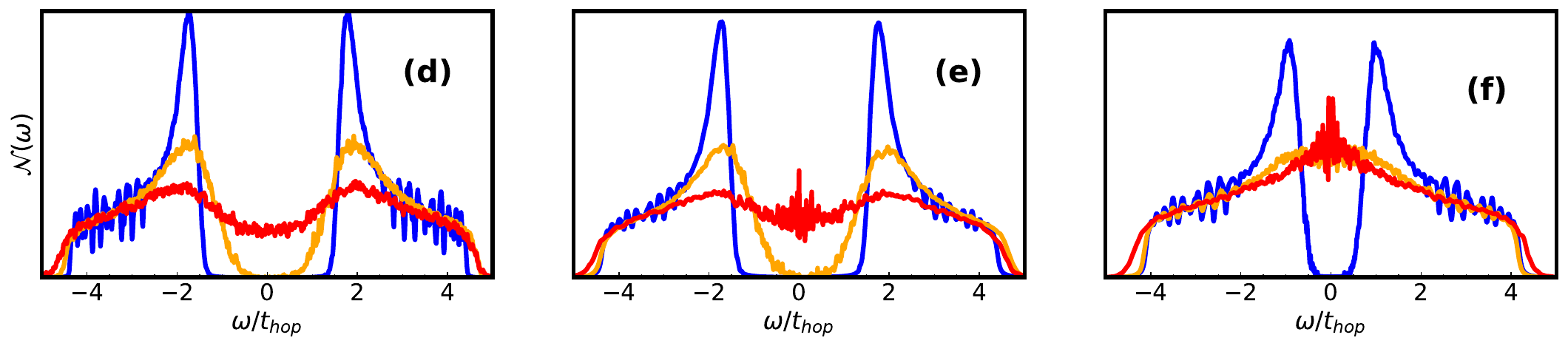}
}
\caption{
Quasi-steady-state lattice displacement distribution and electronic
density of states. The top panels show the distribution of lattice
distortions, $P(x)$, for different values of $T_{\rm bath}$ and
$T_{\rm el}^f$. The bottom panels show the corresponding electronic
density of states computed from the full lattice electronic Hamiltonian
in the final phonon backgrounds.}
\label{fig:dos_steady}
\end{figure*}
%%%%%%%%%%%%%%%%%%%%%%%%%%%%%%%

\subsubsection{Phonon spectrum}
\label{subsubsec:phonon_spectrum}

The phonon dynamical structure factor gives a frequency-resolved 
view of the lattice fluctuations.  We compute $S({\bf q},\omega)$ 
from the space-time Fourier transform of $x_i(t)$.

Fig.\ref{fig:phonon_dsf} shows the spectra for several two-temperature 
conditions.  In equilibrium, the phonon mode near ${\bf Q}=(\pi,\pi)$ 
softens and broadens as the charge-ordering transition is approached.  
The softening indicates a reduced stiffness of the charge-order mode, 
while the broadening reflects enhanced fluctuations and damping.

For a weak pump, $T_{\rm el}/t_{\rm hop}=0.3$, the same qualitative 
behavior occurs, but the critical bath temperature is lower.  The 
excited electronic population has already weakened the ordering 
tendency, so less bath-induced disorder is needed to destroy 
long-range order.  For a strong pump, $T_{\rm el}/t_{\rm hop}=0.7$, 
the spectrum is substantially modified even at low bath temperature. 
 The mode near ${\bf Q}$ is softer and more strongly damped, 
consistent with proximity to amplitude melting.

\subsubsection{Electronic steady state} 
\label{subsubsec:electronic_steady_state} 

We finally characterize the electronic properties of the quasi-steady state. Although the Langevin dynamics is generated using the effective short-range phonon model, the electronic density of states is computed from the full lattice electronic Hamiltonian in the final phonon backgrounds. 
For each late-time configuration $\{x_i\}$, we diagonalize $ h_{ij}=-t_{\rm hop}\delta_{\langle ij\rangle}-g x_i\delta_{ij},$ and average the resulting spectra over time and independent noise realizations. This gives a direct measure of how the local distortions and domain structure affect the electronic gap. Fig.~\ref{fig:dos_steady} shows the late-time lattice displacement distribution $P(x)$ together with the electronic density of states. 

In the charge-ordered regime, $P(x)$ is strongly bimodal, reflecting the two sublattices of the checkerboard state, and the density of states has a clear single-particle gap. Increasing $T_{\rm bath}$ mainly disorders the relative phase of locally distorted regions. As a result, long-range charge order is reduced and in-gap spectral weight appears, but the local distortion distribution can remain bimodal over a substantial range. This is the electronic signature of a polaronic liquid: local lattice distortions survive even after global charge order is weakened. 
Increasing $T_{\rm el}$ produces a qualitatively different effect. The hot electronic population directly weakens the effective ordering potential, reducing the separation between the two peaks in $P(x)$. The charge-order gap is therefore suppressed even when the bath temperature is low. For sufficiently large $T_{\rm el}$, the distribution becomes nearly unimodal and the density of states becomes increasingly metallic. Thus the steady-state electronic spectra support the distinction already suggested by the order parameter and phase diagram: bath heating mainly destroys spatial coherence between locally distorted regions, whereas electronic heating suppresses the local distortion amplitude itself.

\section{Discussion}
\label{sec:discussion}

The main simplification in this work is the treatment of electronic
 relaxation.  The pump-induced electronic population is represented 
by an effective temperature $T_{\rm el}(t)$, while the phonons evolve 
in contact with a bath at $T_{\rm bath}$.  This makes large-lattice, 
long-time simulations possible, but it is not a consistent microscopic 
theory of carrier relaxation.  The electronic sector does not cool 
self-consistently by transferring energy to the phonons, and the 
electronic occupation is not evolved dynamically after 
$T_{\rm el}(t)$ has been specified.  The parameter $T_{\rm el}^f$ 
should therefore be interpreted as a long-lived nonequilibrium
 electronic population, not as a temperature obtained from a 
complete electron-phonon kinetic theory.

This approximation is most appropriate when the charge-order gap 
is large compared with the phonon frequency and bath temperature.  
In that regime, relaxation across the gap is slow and the lattice 
can evolve for an extended time in a quasi-stationary excited
 electronic background.  The model is therefore intended for an 
intermediate-time window: after the pump has generated high-energy 
carriers, but before the electronic population has fully recombined 
and equilibrated with the lattice bath.  The two-temperature 
Langevin scheme should not be viewed as the final stage of
 thermalization; rather, it isolates the effect of a hot electronic
 background on charge-order melting and recovery.

%%%%%%%%%%%%%%%%%%%%%%%%%%%%%%
\begin{figure*}[t]
\centerline{
\includegraphics[width=17cm,height=4cm]{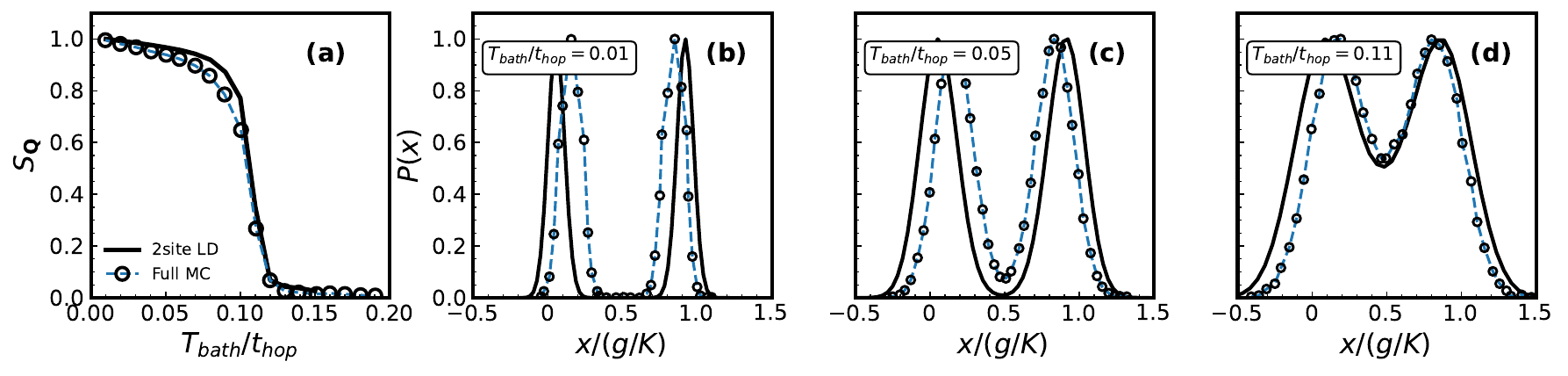}
}
\caption{
Equilibrium benchmarks of the effective 2-site model.
(a) Normalized charge-order structure factor $S_{\bf Q}$
as a function of temperature, compared with adiabatic
Holstein Monte Carlo results.
(b--d) Equilibrium distributions of local distortions
$P(x)$ at representative temperatures. Both the full
Holstein model and the effective Langevin model show
a bimodal distribution at low and intermediate
temperatures, with local distortions persisting close
to $T_{\rm CO}$. This indicates that the loss of charge
order near the transition is mainly associated with
phase/domain melting rather than an immediate collapse
of the local distortion amplitude.
}
\label{fig:benchmark}
\end{figure*}
%%%%%%%%%%%%%%%%%%%%%%%%%%%%%

A more complete theory would allow $T_{\rm el}(t)$ to evolve 
self-consistently.  One possible extension is to couple the 
Langevin dynamics to a rate equation for $T_{\rm el}$, with 
a cooling rate that depends on the instantaneous gap, phonon 
temperature, and distortion amplitude.  Such a scheme could 
capture the feedback in which relaxation is fast when the gap 
is small but becomes bottlenecked as the gap reopens during 
recovery.  Another extension is a three-temperature model with 
separate temperatures for electrons, strongly coupled phonons, 
and the external bath.  A still more microscopic approach 
would evolve the electronic density matrix together with the 
phonons and bath, but this is much more demanding for the 
sizes and times needed to follow domain coarsening.

These extensions would mainly affect the late-time return to 
full equilibrium.  In the present work, the long-time state 
is a quasi-steady state at fixed $T_{\rm el}$ and $T_{\rm bath}$. 
 If the electronic temperature were allowed to relax to the
 bath temperature, the system would eventually move toward the 
equilibrium line of the two-temperature phase diagram.  The 
route to that line would depend on the electronic cooling time.  
Slow cooling allows the system to remain in the two-temperature 
regime long enough for hot carriers to suppress local distortions 
and delay recovery.  Fast cooling would make the dynamics closer
 to ordinary thermal recovery controlled by the phonon bath.  
Thus the missing electronic relaxation does not remove the 
usefulness of the two-temperature phase diagram, but it 
determines how a real pump-probe trajectory moves through it.

\section{Conclusion}
\label{sec:conclusion}

We have introduced a two-temperature Langevin framework for the 
melting and recovery of charge order in the half-filled spinless
Holstein model.  The electronic sector is represented by an 
effective temperature $T_{\rm el}$ 
while the phonons evolve through Langevin dynamics at 
$T_{\rm bath}$.  This gives a tractable semiclassical 
approach to large lattices and long times, where recovery 
involves both local distortion dynamics and collective domain growth.

The transient simulations show clear loss and recovery of charge
order.  Weak pumping suppresses the order parameter without fully
destroying it, leading to relatively smooth recovery.  Strong
pumping can nearly extinguish $S_{\bf Q}$, after which recovery
is slower and more collective.  The recovery time increases with pump
strength and grows rapidly near the pump-dependent
charge-ordering boundary.

The quasi-steady-state phase diagram summarizes the interplay of 
electronic and bath temperatures.  At low $T_{\rm el}$ and low 
$T_{\rm bath}$ the system is charge ordered.  Thermal fluctuations 
produce a disordered polaronic liquid in which local distortions 
remain, while strong electronic heating suppresses the distortion 
amplitude and drives a crossover to a weakly distorted metallic 
regime.  The phonon dynamical structure factor and electronic 
density of states support this picture: the charge-order phonon
 mode softens and broadens near the ordering boundary, and 
the electronic gap is progressively filled and suppressed as 
$T_{\rm el}$ increases.

The main limitation is that electronic relaxation is treated 
phenomenologically.  The effective electronic temperature is 
not obtained from a self-consistent energy-transfer calculation, 
so the final two-temperature state should be interpreted as a 
long-lived intermediate regime rather than the ultimate thermal 
equilibrium state.  Future work should incorporate 
self-consistent electronic cooling, a three-temperature 
description, or direct stochastic mean-field simulations of 
coupled electron-phonon-bath dynamics.  Such extensions would 
determine how real pump-probe trajectories move through the 
two-temperature phase diagram and eventually return to equilibrium.

\vspace{-.5cm}

\acknowledgements{
The authors acknowledge use of the HPC clusters at HRI.  S.S.B. 
was partially supported by the US Department of Energy, 
Basic Energy Sciences under Contract No. DE-SC0020330.
}

\appendix
\section{Benchmarking the two-site approximation}
 The effective model captures the thermal charge-ordering scale 
reasonably well. Its main quantitative error is a larger low-temperature
 value of $S_{\bf Q}$, 
caused by the tendency of the two-site potential to
 overestimate the local distortion amplitude. For the present
 purpose, the important point is that long-range charge order is lost
  in the same temperature range in the two calculations.

The local distortion distribution provides a more microscopic test.
 Fig.\ref{fig:benchmark}  shows P(x) deep in the ordered phase, in the intermediate
  anharmonic regime, and close to the equilibrium transition. Both
  methods show a clear bimodal distribution at low temperature,
  corresponding to the two sublattices of the checkerboard charge order.
   With increasing temperature the peaks broaden and overlap,
   but the distribution remains bimodal near the transition.

This behavior is characteristic of strong-coupling charge-order
melting. Local distortions remain finite near and above the ordering
 temperature, while long-range coherence between charge-ordered
  regions is lost. Equilibrium melting is therefore primarily
  phase/domain driven. This benchmark sets the reference for the
  nonequilibrium results, where increasing $T_{\rm el}$ produces a
   more direct suppression of the distortion amplitude.

\bibliography{copp}

\end{document}